\documentclass[prd,twocolumn]{revtex4}%
\usepackage{epsfig}
\usepackage{amsmath}
\usepackage{verbatim}
\usepackage{graphicx}%
\usepackage{amsfonts}%
\usepackage{amssymb}

\begin{document}

\title{Topological charge screening and pseudoscalar glueballs}
\author{Hilmar Forkel}
\affiliation{IFT - Universidade Estadual Paulista, Rua Pamplona, 145, 01405-900 Sao Paulo,
SP, Brazil }
\affiliation{Institut f\"{u}r Theoretische Physik, Universit\"{a}t Heidelberg, D-69120
Heidelberg, Germany }

\begin{abstract}
Topological charge screening in the QCD vacuum is found to provide crucial
nonperturbative contributions to the short-distance expansion of the
pseudoscalar ($0^{-+}$) glueball correlator. The screening contributions enter
the Wilson coefficients and are an indispensable complement to the direct
instanton contributions. They restore consistency with the anomalous axial
Ward identity and remedy several flaws in the $0^{-+}$ glueball sum rules
caused by direct instantons in the absence of screening (lack of resonance
signals, violation of the positivity bound and of the underlying low-energy
theorem). The impact of realistic instanton size distributions and the
(gauge-invariant) renormalization of the instanton contributions are also
discussed. New predictions for the $0^{-+}$ glueball mass and decay constant
are presented.

\end{abstract}
\maketitle

\section{Introduction}

The glueball sector of QCD has remained intriguing and challenging since the
early days of QCD \cite{gel72} and  glueball physics offers
promising  opportunities for the study of low-energy gluon dynamics and of the
often elusive gluonic component in hadron wavefunctionals. One such
opportunity was recently exploited in OCD sum rule analyses, which found
nonperturbative short-distance physics in the form of direct instantons
\cite{inst} to play a crucial role in the structure and dynamics of the scalar
($0^{++}$) glueball \cite{for01,for03}. Indeed, the instanton-improved
operator product expansion (IOPE) of the $0^{++}$ glueball correlator resolves
two longstanding problems of the conventional sum rules (the mutual
inconsistency of different Borel moment sum rules and the conflict with the
underlying low-energy theorem \cite{nov280,nar98}), generates new scaling
relations between fundamental glueball and instanton properties, and leads to
improved sum rule predictions for scalar glueball properties \cite{for01}.
(See also the subsequent gaussian sum rule analysis \cite{har01}, based on the
same instanton contributions.) The Borel sum rule analysis has recently been
improved and extended (to realistic instanton size distributions and
renormalized instanton contributions) in \cite{for03}. Hard
nonperturbative contributions to the pseudoscalar glueball IOPE and sum rules
were also investigated. On these we will report in the following. A
comprehensive analysis of both spin-0 glueball channels can be found in Ref.
\cite{for03}. 

The implementation of the direct instanton contributions in the $0^{-+}$
channel is straightforward: the expressions from the $0^{++}$ channel
\cite{for01} can simply be taken over, with their signs inverted. This is a
consequence of the (Minkowski) (anti-) self-duality of the (anti-) instanton's
field strength, $G_{\mu\nu}^{\left(  I,\bar{I}\right)  }=\pm i\tilde{G}%
_{\mu\nu}^{\left(  I,\bar{I}\right)  }$. However, one immediately suspects
that adding the dominant and, due to the sign change, strongly repulsive
instanton contributions will seriously unbalance the Borel sum rules. Two of
the consequences were recently observed in \cite{zha03}: any reliable signal
for a pseudoscalar glueball resonance disappears (in contradiction to lattice
evidence), and even the fundamental spectral positivity bound is violated. In
addition, the crucial low-energy theorem for the zero-momentum $0^{-+}$
glueball correlator, and therefore the underlying anomalous axial Ward
identity, is strongly violated \cite{for03}. As pointed out in \cite{for03},
these problems have an appealing solution in the form of additional
nonperturbative short-range contributions to the IOPE, associated with
topological charge screening. Below, we will sketch the implementation of the
screening contributions and their impact on the sum rule analysis. We also
comment on the effects of realistic instanton size distributions and the
renormalization of the instanton-induced Wilson coefficients.

\section{Correlator, IOPE and sum rules}

Our discussion will be based on the pseudoscalar glueball correlation function%
\begin{equation}
\Pi_{P}\left(  x\right)  =\left\langle 0|T\,O_{P}\left(  x\right)
O_{P}\left(  0\right)  |0\right\rangle ,
\end{equation}
where $O_{P}$ is the standard gluonic interpolating field
\begin{equation}
O_{P}\left(  x\right)  =\alpha_{s}G_{\mu\nu}^{a}\left(  x\right)  \tilde
{G}^{a\mu\nu}\left(  x\right)  \label{tipf}%
\end{equation}
($\tilde{G}_{\mu\nu}\equiv\left(  i/2\right)  \varepsilon_{\mu\nu\rho\sigma
}G_{\rho\sigma}$), and its Fourier transform%
\begin{equation}
\Pi_{P}(-q^{2})=i\int d^{4}x\,e^{iqx}\left\langle 0|T\,O_{P}\left(  x\right)
O_{P}\left(  0\right)  |0\right\rangle .
\end{equation}
The zero-momentum limit of this correlator is governed by the low-energy
theorem (LET) \cite{pslet}
\begin{equation}
\Pi_{P}\left(  q^{2}=0\right)  =\left(  8\pi\right)  ^{2}\frac{m_{u}m_{d}%
}{m_{u}+m_{d}}\left\langle \bar{q}q\right\rangle \label{pLET}%
\end{equation}
(for three light flavors and $m_{u,d}\ll m_{s}$) which derives from the axial
anomaly and imposes stringent consistency requirements on the sum rule
analysis  \cite{for03}.

The hadronic information in the glueball correlators is most directly
accessible in the dispersive representation%
\begin{equation}
\Pi_{P}\left(  Q^{2}\right)  =\frac{1}{\pi}\int_{s_{i}}^{\infty}%
ds\frac{\operatorname{Im}\Pi_{P}\left(  -s\right)  }{s+Q^{2}}\label{disprel}%
\end{equation}
where the necessary number of subtractions is implied. The QCD sum-rule
description of the spectral functions contains one or two resonances (in
zero-width approximation) and the local-duality continuum, i.e.
\begin{align}
\operatorname{Im}\Pi_{P}^{\left(  ph\right)  }\left(  s\right)   &  =\pi
\sum_{i=1}^{2}f_{Pi}^{2}m_{Pi}^{4}\delta\left(  s-m_{Pi}^{2}\right)
\label{phspecdens} \nonumber \\
&  +\theta\left(  s-s_{0}\right)  \operatorname{Im}\Pi_{P}^{\left(
IOPE\right)  }\left(  s\right)
\end{align}
The continuum representation covers the invariant-mass region ''dual'' to
higher-lying resonances and multi-hadron continuum, starting at an effective
threshold $s_{0}$. It is obtained from the discontinuities of the IOPE, i.e.
the expansion of the correlator at large, spacelike momenta $Q^{2}\equiv
-q^{2}\gg\Lambda_{QCD}$ into condensates of operators $\hat{O}_{D}$ of
increasing dimension $D$, \
\begin{equation}
\Pi_{P}(Q^{2})=\sum_{D=0,4,...}\tilde{C}_{D}^{\left(  P\right)  }\left(
Q^{2};\mu\right)  \left\langle 0\left|  \hat{O}_{D}\right|  0\right\rangle
_{\mu}\text{ }.
\end{equation}
``Hard'' field modes with momenta $\left|  k\right|  >\mu$ contribute to the
momentum-dependent Wilson coefficients $\tilde{C}_{D}\left(  Q^{2}\right)  $
while ``soft'' modes with $\left|  k\right|  \leq\mu$ generate the condensates
$\left\langle \hat{O}_{D}\right\rangle $.

In order to write down the sum rules, the IOPE - with the continuum subtracted
and weighted by powers of $-Q^{2}$ - is Borel-transformed,%
\begin{align}
{\cal R}_{P,k}^{\left(  IOPE\right)  }\left(  \tau;s_{0}\right)   &
=\hat{B}\left[  \frac{\left(  -Q^{2}\right)  ^{k}}{\pi}\int_{0}^{s_{0}%
}ds\frac{\operatorname{Im}\Pi_{P}^{\left(  IOPE\right)  }\left(  -s\right)
}{s+Q^{2}}\right]  \\ \nonumber 
&  =-\delta_{k,-1}\Pi_{P}^{\left(  IOPE\right)  }\left(  0\right)  \\ \nonumber
&  +\frac{1}{\pi}\int_{0}^{s_{0}}dss^{k}\operatorname{Im}\Pi_{P}^{\left(
IOPE\right)  }\left(  s\right)  e^{-s\tau},
\end{align}
for $k\geq-1$. The hadronic parameters $m_{Pi},f_{Pi},s_{0}$ are then
determined by matching these moments in the fiducial $\tau$-region\ to their
resonance-induced counterparts (and a subtraction constant for $k=-1$, fixed
by the LET (\ref{pLET})). The resulting IOPE sum rules are%
\begin{equation}
{\cal R}_{P,k}^{\left(  IOPE\right)  }\left(  \tau;s_{0}\right)
=\sum_{i=1}^{2}f_{Pi}^{2}m_{Pi}^{4+2k}e^{-m_{Pi}^{2}\tau}-\delta_{k,-1}\Pi
_{P}^{\left(  ph\right)  }(0).\label{rsrs}%
\end{equation}

The perturbative contributions to the IOPE coefficients, i.e. the conventional
OPE,\ can be found in \cite{nov279,asn92,zha03,for03}. In the following we
will focus on the nonperturbative contributions due to direct instantons and
topological charge screening.

\section{Direct instantons}

Dominant direct instanton contributions to the $0^{-+}$ glueball\ correlator
are received by the unit-operator IOPE coefficient, $\tilde{C}_{0}^{\left(
P,I+\bar{I}\right)_D}=\Pi_{P}^{\left(  I+\bar{I}\right) }$. They are best
calculated in $x$-space, with the result
\begin{equation}
\tilde{C}_{0}^{\left(P,I+\bar{I}\right)_D}\left(  x^{2}\right)
= - \frac{2^{8}  3}{7}\int d\rho n\left(  \rho\right)  \frac{1}{\rho^{4}}%
\,_{2}F_{1}\left(  4,6,\frac{9}{2},-\frac{x^{2}}{4\rho^{2}}\right)
.\label{dirincof}%
\end{equation}
The further evaluation requires the (anti-) instanton distribution
$n_{I,\bar{I}}\left(  \rho\right)  $ with its two leading moments, the
instanton density $\bar{n}$ and average size $\bar{\rho}$, as input. All
previous studies of direct instanton effects have relied on the simplest
possible, spike-like approximation $n(\rho)=\bar{n}\delta\left(  \rho
-\bar{\rho}\right)  $. In Ref.\cite{for03} a realistic
finite-width distribution was
implemented instead, which is fully determined
by $\bar{n}$, $\bar{\rho}$ and the known small- and large-$\rho$ behavior
\cite{rin99},  (for $N_{c}=N_{f}=3$)
\begin{equation}
n_{g}\left(  \rho\right)  =\frac{2^{18}}{3^{6}\pi^{3}}\frac{\bar{n}}{\bar
{\rho}}\left(  \frac{\rho}{\bar{\rho}}\right)  ^{4}\exp\left(  -\frac{2^{6}%
}{3^{2}\pi}\frac{\rho^{2}}{\bar{\rho}^{2}}\right)  .
\end{equation}

From the Fourier transform of (\ref{dirincof}) one finds the
direct-instanton-induced Borel moments \cite{for01}
\begin{align}
{\cal L}_{k-1}^{\left(  I+\bar{I}\right)  }\left(  \tau\right)   &
=-2^{6}\pi^{2}\frac{-\partial^{k}}{\partial\tau^{k}}\int d\rho n\left(
\rho\right)  x^{2}e^{-x} \nonumber \\
&  \times\left[  \left(  1+x\right)  K_{0}\left(  x\right)  +\left(
2+x+\frac{2}{x}\right)  K_{1}\left(  x\right)  \right]  ,
\end{align}
($x=\rho^{2}/2\tau,$ $k\geq0$) and from the imaginary part \cite{for01}%
\begin{equation}
\operatorname{Im}\Pi_{P}^{\left(  I+\bar{I}\right)  }\left(  -s\right)
=-2^{4}\pi^{4}\int d\rho n\left(  \rho\right)  \rho^{4}s^{2}J_{2}\left(
\sqrt{s}\rho\right)  Y_{2}\left(  \sqrt{s}\rho\right)  \label{impinsts}%
\end{equation}
at timelike momenta one then has%
\begin{align}
{\cal R}_{P,k}^{\left(  I+\bar{I}\right)  }\left(  \tau\right)   &
=-2^{7}\pi^{2}\delta_{k,-1}\int d\rho n\left(  \rho\right)  -2^{4}\pi^{3}\int
d\rho \nonumber \\
&  \times n\left(  \rho\right)  \rho^{4}\int_{0}^{s_{0}}dss^{k+2}J_{2}\left(
\sqrt{s}\rho\right)  Y_{2}\left(  \sqrt{s}\rho\right)  e^{-s\tau}.
\end{align}
The direct-instanton contributions (with subtracted continuum) are
an important complement to  the corresponding perturbative ones.

The realistic (finite-width) instanton size distribution tames the rising
oscillations of the imaginary part (\ref{impinsts}) at large $s$ (an artefact
with misleading impact on the $s_{0}$ dependence of the moments) into a strong
decay $\propto s^{-5/2}$. Furthermore, it allows for a gauge-invariant
renormalization of the instanton-induced coefficients\ by excluding
contributions from instantons with size $\rho>\mu^{-1}$, i.e. by replacing
\begin{equation}
n\left(  \rho\right)  \rightarrow\tilde{n}_{\mu}\left(  \rho\right)
\equiv\theta_{\beta}\left(  \rho-\mu^{-1}\right)  n\left(  \rho\right)
\end{equation}
with a ``soft'' step function $\theta_{\beta}$ (e.g. $\theta_{\beta}\left(
\rho-\mu^{-1}\right)  =\left[  \exp\left(  \beta\left(  \rho-\mu^{-1}\right)
\right)  +1\right]  ^{-1}$). The instanton-induced $\mu$-dependence turns out
to be relatively weak for $\mu<\bar{\rho}^{-1}$, in complicance with the other
sources of $\mu$ dependence. Note that the standard spike distribution (with
$\bar{\rho}<\mu^{-1}$) misses the reduction of the total  instanton density to
the direct instanton part,%
\begin{equation}
\bar{n}=\int_{0}^{\infty}d\rho n\left(  \rho\right)  \rightarrow\int
_{0}^{\infty}d\rho\tilde{n}_{\mu}\left(  \rho\right)  \equiv\bar{n}%
_{direct}.\label{ndir}%
\end{equation}

\section{Topological charge screening}

As argued above, it is not surprising that the dominant and strongly repulsive
direct instanton contributions, when added as the \textit{sole}
nonperturbative contributions to the IOPE coefficients, upset the $0^{-+}$
glueball sum rules and have the mentioned, detrimental effects. In fact, this
suggests that additional  important contributions, which should predominantly
affect the pseudoscalar IOPE, are still amiss. And indeed, the $0^{-+}$
glueball correlator is proportional to the topological charge correlators and
therefore maximally sensitive to the short-distance topological charge
(probably mainly instanton - antiinstanton) correlations in the QCD vacuum.
Their impact on the pseudoscalar glueball correlators was found to be
exceptionally strong in the instanton liquid model \cite{inst, sch95}.
Topological charge correlations are created by light-quark loops or,
equivalently at low energies, by the attractive (repulsive) $\eta^{\prime}%
$-meson exchange forces between opposite-sign (equal-sign) topological
charges. They lead to Debye screening of the topological charge with
``screening mass'' $m_{\eta^{\prime}}$ and the corresponding (small!)
screening length $\lambda_{D}\sim m_{\eta^{\prime}}^{-1}\sim0.2$ fm
\cite{dow92,kik92}. Since $m_{\eta^{\prime}}>\mu$, the screening correlations
contribute to the Wilson coefficients of the pseudoscalar glueball correlator.

The screening contributions can be obtained from the coupling of the
$\eta^{\prime}$ mesons to the topological charge density, as dictated by the
axial anomaly \cite{div80}. In the chiral limit, where $\eta^{\prime}=\eta
_{0}$ is purely flavor-singlet, and for $k\lesssim m_{\eta^{\prime}}$ the
coupling to the topological charges in the vacuum medium (approximated for
simplicity as concentrated in pointlike instantons) is governed by the
effective lagrangian \cite{dow92,kik92}%
\begin{equation}
{\cal L}=\frac{1}{2}\left(  \partial\eta_{0}\right)  ^{2}-2\bar{n}%
\cos\left(  \gamma_{\eta_{0}}\eta_{0}+\theta\right)
\end{equation}
where $\bar{n}$ is the global topological charge density and where we have
introduced a source $\theta\left(  x\right)  $ for the local topological
charge density $Q\left(  x\right)  $. Note the screening mass $m_{scr}%
^{2}=2\bar{n}\gamma_{\eta_{0}}^{2}$ of the $\eta_{0}$. Taking two derivatives
of the corresponding generating functional with respect to $\theta$ leads (for
small amplitudes $\eta_{0}$) to the topological charge correlator
\begin{align}
\left\langle Q\left(  x\right)  Q\left(  0\right)  \right\rangle  &  =\Pi
_{P}\left(  x\right)  /\left(  8\pi\right)  ^{2} \nonumber \\
&  \simeq-2\bar{n}\delta^{4}\left(  x\right)  -\left(  2\bar{n}\gamma
_{\eta_{0}}\right)  ^{2}\left\langle \eta_{0}\left(  x\right)  \eta_{0}\left(
0\right)  \right\rangle .
\end{align}
The first term is just the pointlike approximation to the direct-instanton
contribution evaluated above. The second one is the screening correction,
which modifies the nonperturbative contributions to the pseudoscalar IOPE
coefficient into%
\begin{equation}
\Pi_{P}\left(  x\right)  =\Pi_{P}^{\left(  I+\bar{I}\right)  }\left(
x\right)  -\left(  \bar{n}\gamma_{\eta_{0}}\right)  ^{2}\frac{m_{\eta_{0}}%
}{\pi^{2}x}K_{1}\left(  m_{\eta_{0}}x\right)  .\label{pinandscr}%
\end{equation}
After implementing finite quark-mass effects (i.e. $\eta_{0}$-$\eta_{8}$
mixing), the (Minkowski) Borel moments associated with the screening
contributions become
\begin{align}
{\cal R}_{P,k}^{\left(  scr\right)  }\left(  \tau\right)   &
=-\delta_{k,-1}\left(  \frac{F_{\eta^{\prime}}^{2}}{m_{\eta^{\prime}}^{2}%
}+\frac{F_{\eta}^{2}}{m_{\eta}^{2}}\right)  \label{rkscr} \nonumber \\
&  +F_{\eta^{\prime}}^{2}m_{\eta^{\prime}}^{2k}e^{-m_{\eta^{\prime}}^{2}\tau
}+F_{\eta}^{2}m_{\eta}^{2k}e^{-m_{\eta}^{2}\tau}.
\end{align}

The $\tau$-independent term in Eq. (\ref{rkscr}) is the screening-induced
subtraction constant $-\Pi_{P}^{\left(  scr\right)  }\left(  0\right)  $. It
shows that inclusion of the screening contributions is mandatory in order to
satisfy the axial $U\left(  1\right)  $ Ward identity and the ensuing LET
(\ref{pLET}). Indeed, direct instantons generate a large subtraction constant
$\Pi_{P}^{\left(  I+\bar{I}\right)  }\left(  0\right)  =-2^{7}\pi^{2}\bar{n}$
while the LET demands the zero-momentum limit of the physical correlator to be
of the order of the light quark masses, i.e. much smaller. The screening
contribution (\ref{rkscr}) is necessary to cancel most of it and to restore
consistency with the LET. In the chiral limit, the screening contribution
turns into
\begin{equation}
\Pi_{P}^{\left(  scr\right)  }\left(  0\right)  =\frac{F_{\eta^{\prime}}^{2}%
}{m_{\eta^{\prime}}^{2}}=\frac{\left(  16\pi\bar{n}\gamma_{\eta_{0}}\right)
^{2}}{2\bar{n}\gamma_{\eta_{0}}^{2}}=2^{7}\pi^{2}\bar{n}\label{psubtrscr}%
\end{equation}
and the cancellation becomes exact (due to the infinite-range interactions
mediated by massless Goldstone bosons). The above argument provides compelling
evidence for the screening contributions to be an indispensable complement to
the direct instantons.

The cancellation between the subtraction terms suggests a simple strategy for
renomalizing the screening contributions. Since the large-$\rho$ cutoff
$\mu^{-1}$ amounts to replacing $\bar{n}$ by $\bar{n}_{dir}=\zeta\bar{n}$ with
$\zeta<1$ (cf. Eq. (\ref{ndir})), consistency with the LET requires the same
replacement in the screening contributions (\ref{pinandscr}). 

Besides restoring the axial Ward identity, inclusion of the screening
contributions resolves the positivity-bound violation and creates a strong
signal for both the $\eta^{\prime}$ and the pseudoscalar glueball resonances
in the corresponding Borel sum rules. The screening contributions
(\ref{rkscr}) are of substantial size, even relative to the direct-instanton
contributions, and they are largest at small and intermediate $\tau
\lesssim\lambda_{scr}^{2}\sim1$ GeV$^{2}$. Moreover, they modify qualitative
features of the Borel moments (e.g. the sign of the slope) to which the sum
rule fits are very sensitive. Hence, trustworthy sum rule results cannot be
obtained even at small and intermediate $\tau$ when the screening
contributions are ignored \cite{zha03}. Even  the bound obtained in
\cite{zha03} has therefore to be regarded as invalid. 

After including the screening contributions, all four Borel-moment sum rules
(\ref{rsrs}) are stable and yield consistent results. Note that previous
analyses of the $0^{-+}$ sum rules have discarded the $k=-1$ sum rule and
therefore missed the chance to implement the first-principle information from
the low-energy theorem, as well as a very useful consistency check. The
two-resonance fit is clearly favored over the one-pole approximation, i.e. the
IOPE provides clear signals for the $\eta^{\prime}$ resonance (with small
$\eta$ admixtures due to mixing) in addition to a considerably heavier
$0^{-+}$ glueball. The $0^{-+}$ glueball mass is found to be $m_{P}%
=2.2\pm 0.2$ GeV, and the decay constant $f_{P}=0.6\pm0.25$ GeV. (For a
complete discussion see \cite{for03}.)

\section{Summary}

We have reported recent developments in understanding and evaluating
nonperturbative glueball physics in the pseudoscalar channel on the basis of
the operator product expansion \cite{for03}. Contrary to naive expectation,
much of this nonperturbative physics takes place at surprisingly short
distances $\left|  x\right|  \sim0.2-0.3$ fm, and consequently shows up in the
Wilson coefficients of the IOPE. Direct instantons are the paradigm for such
physics, and their contributions to the spin-0 glueball correlators are indeed
exceptionally large. We have improved on the previous evaluation of these
contributions \cite{for01} by implementing a realistic instanton size
distribution and the renomalization of the instanton-induced coefficients
(both of these improvements  should be useful in other hadron channels as well). A further new
development is very specific to the $0^{-+}$ glueball channel: we have found
compelling evidence for topological charge screening to provide crucial
contributions to the unit operator coefficient of the pseudoscalar glueball
IOPE. The screening contributions form an indispensable complement to the
direct instantons, roughly speaking ``unquenching'' them and thereby restoring
the axial Ward identity. Moreover, they balance the strong repulsion of the
direct instantons, correct the otherwise gross violation of the LET, resolve
the violation of the spectral positivity bound and generate a strong (and
otherwise absent) signal for the $0^{-+}$ glueball resonance. With screening
included, all Borel moment sum rules provide consistent and stable predictions
for the fundamental $0^{+-}$ glueball properties ($m_{P}=2.2\pm0.2$ GeV,
$f_{P}=0.6\pm0.25$ GeV).

This work was supported by FAPESP.

\end{document}